\documentclass[preprint,aps,prl,preprintnumbers,amsmath,amssymb,superscriptaddress]{revtex4}
\usepackage[a4paper, total={6.5in, 9in}]{geometry}
\usepackage[colorlinks=true,linkcolor=black, citecolor=blue, urlcolor=blue, 
    unicode=true]{hyperref}
    
\usepackage{graphicx}
\usepackage{amsmath}
\usepackage{blindtext}
\usepackage[version=4]{mhchem}
\usepackage{hyperref}
\usepackage{color}
\usepackage{ulem}
\usepackage{times}
\usepackage{soul}
\usepackage{afterpage}

\def\Revise{\color[rgb]{0,0,1}}

\newcounter{lastnote}

\begin{document}
\title{Non-equilibrium Effects in Vibrational Modes Pumped by Inelastic Tunneling}
\author{Chi Ming Yim}
\email{c.m.yim@sjtu.edu.cn}
\affiliation{State Key Laboratory of Micro-nano Engineering Science, Tsung-Dao Lee Institute \& School of Physics and Astronomy, Shanghai Jiao Tong University, Shanghai 201210, China}
\affiliation{SUPA, School of Physics and Astronomy, University of St Andrews, North Haugh, St Andrews, Fife, KY16 9SS, United Kingdom}
\author{Owain T. Beynon}
\affiliation{Department of Physics, King’s College London, London WC2R 2LS, United Kingdom}
\author{Seunghyun Khim}
\affiliation{Max Planck Institute for Chemical Physics of Solids, N\"othnitzer Stra\ss e 40, 01187 Dresden, Germany}
\author{Chiara Gattinoni}
\affiliation{Department of Physics, King’s College London, London WC2R 2LS, United Kingdom}
\author{Peter Wahl}
\affiliation{SUPA, School of Physics and Astronomy, University of St Andrews, North Haugh, St Andrews, Fife, KY16 9SS, United Kingdom}
\affiliation{Physikalisches Institut, Universität Bonn, Nussallee 12, 53115 Bonn, Germany}
\date{\today}

\begin{abstract} 
The properties of strongly correlated electron materials exhibit a surprising sensitivity to small lattice distortions, providing an opportunity for their tuning by selective distortion driving, usually achieved by optical excitations. Using inelastic electron tunneling in scanning tunneling microscopy, we demonstrate that at the surface of a strongly correlated electron material, we can drive vibrational excitations out of equilibrium, by studying the dynamics of localized modes on the \ce{Pd}-terminated surface of the delafossite oxide \ce{PdCrO2}. This surface forms a tiling of hydrogen clusters of varying sizes and shapes upon hydrogen adsorption. Our findings reveal that vibrational excitations in the clusters exhibit longer lifetimes than on typical metal surfaces. Detailed analysis of the spectroscopy data reveals signatures of non-equilibrium effects in the excitations which we attribute to the extended lifetimes of these modes. Theoretical calculations support that the long-lived nature of the excitations is related to the unique properties of the substrate.
\end{abstract}

\maketitle

\section{INTRODUCTION}
Vibrations are fundamental excitations of matter associated with lattice distortions. In molecules, vibrations occur when two neighboring atoms oscillate in opposite directions, leading to bond distortions at characteristic frequencies. In solids, the situation is more complex: the displacement of one atom/ion also leads to displacement of its neighbors, resulting in collective motions of all atoms/ions. In crystalline solids, such collective motions follow phonon-dispersion relations, with typical phononic bandwidths ranging from a few $\mathrm{meV}$ to tens of $\mathrm{meV}$. The interaction of lattice vibrations and distortions with the electrons in solids drives a wide range of phenomena from charge density waves to superconductivity. This intricate coupling implies that driving specific vibrational modes promises control over electronic ground states \cite{cavalleri_photo-induced_2018}.

Vibration dynamics in solids has been widely studied using techniques including Raman and infrared absorption spectroscopy.  Relevant studies performed at the atomistic level, especially those focusing on driving localized modes in quantum materials, remain rare.

Inelastic electron tunneling spectroscopy (IETS) combined with the atomic resolution of scanning tunneling microscopy (STM) is a powerful technique for mapping inelastic processes on surfaces, including vibrations of adsorbates \cite{stipe_single-molecule_1998}, spin excitations \cite{madhavan_tunneling_1998,hirjibehedin_spin_2006}, and surface phonon excitations \cite{minamitani_surface_2016}. Recently, vibration signals have also been used to probe strongly correlated many-body systems in molecular overlayers \cite{eickhoff_inelastic_2020} and map the local potential energy landscapes of molecules \cite{chiang_real-space_2014}.

To study the non-equilibrium excitation of a vibrational mode on a strongly-correlated electron material using STM-IETS, here we chose the \ce{Pd} terminated surface of the delafossite oxide \ce{PdCrO2} with chemisorbed hydrogen. Delafossite oxides \ce{ABO2} host a wide variety of exciting properties due to their innate layered structure, the potential for property tuning by careful selection of the A and B cations \cite{cheong_multiferroics_2007,kawazoe_p-type_1997,singh_electronic_2007} and the availability of samples of exceptionally high quality. In particular, delafossite oxides \ce{PdCoO2} and \ce{PdCrO2} have recently drawn much attention due to the extremely long mean free path of the conduction electrons \cite{hicks_quantum_2012}, and as promising candidates for the realization of ballistic transport in microstructures \cite{bachmann_directional_2022}. The surfaces of delafossites are polar because of the charge distribution within the unit cell: in \ce{PdCrO2}, the \ce{Pd} ions carry a nominal charge of $+1$ and the \ce{CrO2} layers of $-1$ [Fig.~\ref{fig-intro}(a)] \cite{mackenzie_properties_2017}. Their polar nature leads to surface properties drastically different from the bulk \cite{sunko_maximal_2017,yim_quasiparticle_2021,yim_avoided_2024,mazzola_itinerant_2018,mazzola_tuneable_2022}.

Recently, we found an aperiodic tiling structure formed by hydrogen on the \ce{Pd}-terminated surface of \ce{PdCrO2} \cite{yim_adsorbate-induced_2025}. Comprised of tightly packed clusters of hydrogen of different sizes and shapes, the tiling structure possesses no translational or rotational symmetries [see Figs.~\ref{fig-intro}(a) and 1(b) for its schematic model].  Notably, its formation also leads to sharp inelastic vibrational features in inelastic tunneling spectra recorded within the clusters. Therefore, it is an ideal platform to address the above question.

Using IETS in STM to characterize these inelastic features, our data show that they have extremely narrow line widths and demonstrate a non-equilibrium effect in inelastic tunneling as a result of driving of the adsorbate-substrate system into an excited state. Fitting the data using a two-level excitation model, we find that the vibrational modes of the Pd-bonded H have longer lifetimes than would be expected for a metal surface, reaching hundreds of picoseconds.

\section{II. RESULTS AND DISCUSSION}
\subsection{Hydrogen tiling on \ce{Pd/PdCrO2}}
The ways of distinguishing between the \ce{Pd} and \ce{CrO2} terminated surfaces of \ce{PdCrO2} in STM have been described elsewhere \cite{yim_avoided_2024,yim_adsorbate-induced_2025}, so here we focus only on the \ce{Pd}-terminated surface.  The \ce{Pd}-terminated surface in its pristine form appears extremely flat and exhibits Friedel oscillations from defects [Fig.~\ref{fig-intro}(c)] \cite{mazzola_tuneable_2022,yim_adsorbate-induced_2025,Zheng_2026}.  Adsorption of hydrogen on this surface results in a tiling structure comprising tightly packed clusters of hydrogen atoms bonded to the Pd surface layer [Fig.~\ref{fig-intro}(d)], leading to strong variations of the local surface work function and density of states throughout the tiling and electronic localization within the clusters \cite{yim_adsorbate-induced_2025}.

Another important consequence of its formation is the inelastic excitation features arising from the vibrations of the H-Pd bonds formed within the clusters. To demonstrate this, in Fig.~\ref{fig-inelastic-overall}(a) we show a differential conductance $g(V)$ spectrum recorded in the center of one of the clusters in the nonperiodic tiling, showing a number of step-like features (marked by arrows) at specific energies.  In the corresponding $g'(V)$ spectrum [Fig.~\ref{fig-inelastic-overall}(b)], the step-like features appear as peak-dip pairs at the same energy at both positive and negative bias voltages.  One key signature of inelastic contributions to the differential conductance is their symmetry across zero bias.  On this ground, and to confirm the symmetry of the inelastic features across zero bias, we show the $g'(V)$ spectrum with its $-g'(-V)$ counterpart in Fig.~\ref{fig-inelastic-overall}(b). From the overlap of the spectra, we identify six features that can be clearly assigned to inelastic contributions: peak $\alpha$ at $3~\mathrm{mV}$, $\beta$ at $15~\mathrm{mV}$, peaks $\gamma_1$ and $\gamma_2$ at $42~\mathrm{mV}$ and $84~\mathrm{mV}$, peak $\delta$ at $59~\mathrm{mV}$, and one sharp peak $\varepsilon$ at $274~\mathrm{mV}$.

\subsection{Assignment of the inelastic features}
Our previous work has shown that the inelastic peak $\gamma_1$ ($42~\mathrm{meV}$) corresponds to the vibration mode along the in-plane directions of the \ce{H}-\ce{Pd} bonds, and $\varepsilon$ ($274~\mathrm{meV}$) along the out-of-plane direction. $\gamma_2$ ($84~\mathrm{meV}$), with a frequency twice of $\gamma_1$, is the second harmonic of the former. The presence of higher harmonic peaks in STM-IETS signifies higher-order excitations \cite{Hipps_Inelastic_1993,Lykkebo_Strong_2021}, typically arising from phenomena such as vibronic coupling \cite{Qiu_Vibronic_2004,Reecht_Vibrational_2020,Wang_Topologically_2024}, overtone excitations \cite{Czap_detection_2019}, and nonequilibrium pumping \cite{Frederiksen_Theory_2014}. Given the simple geometry of our setup, we attribute $\gamma_2$ to overtone excitation. Its presence suggests progressive pumping of the vibrational mode.

To determine the origins of other inelastic peaks, we look for normal modes of vibration present within the substrate \ce{PdCrO2}.  By comparison of our data with the calculations by Kumar \textit{et al.} \cite{kumar_first_2013}, we find that the energy of the peak $\beta$ ($15~\mathrm{meV}$) is comparable to that of the $\mathrm{E_{u}^{\downarrow}}$ mode and of $\delta$ ($59~\mathrm{meV}$) to the $\mathrm{E_{g}}$ mode. We cannot associate $\alpha$ ($3~\mathrm{mV}$) with a vibrational mode, suggesting that it is of a different origin. More details on the peak assignment can be found in Supplemental Note I in Supplemental Material \cite{SM_2025}.

\subsection{Characteristics of the in-plane vibration mode}
Due to the structural complexity of the non-periodic tiling, we expect all vibrational modes to vary spatially.  To verify this, we performed spectroscopic imaging measurements of $g'(\mathbf{r}, V)$ in a $(8\mathrm{nm})^2$ surface region comprising $\sim 50$ hydrogen clusters of different sizes and shapes [Fig.~\ref{fig-inelastic-overall}(a)], to spatially map the energy and intensity of inelastic excitations.  From the topographic image [Fig.~\ref{fig-inelastic-overall}(c)] and $g'(\mathbf{r}, V)$ map slices recorded on and off the peak energy of the $\gamma_1$ mode [Figs.~\ref{fig-inelastic-overall}(d) and \ref{fig-inelastic-overall}(e)], we observe that $\gamma_1$ is strongest in the center of each cluster and becomes significantly weaker in intensity in the boundary region [also see the averaged spectra extracted from those two positions shown in Fig.~\ref{fig-inelastic-overall}(f)]. Within the clusters, the intensity is strongest in the center and decays towards the edge (see Supplemental Note II and Supplemental Fig. S1 to S5). 

The inelastic features are all extremely sharp: Unlike inelastic modes arising from excitation of a delocalized phonon mode that has an energy width comparable to its characteristic band width, the observed inelastic feature $\gamma_1$ has a width in energy that is limited by the resolution of the instrument, i.e. thermal broadening, rather than by an intrinsic lifetime of the corresponding bosonic mode  [Fig.~\ref{fig-inelastic-analysis}(a)]. This suggests the highly localized nature of the observed $\gamma_1$ mode.

The $\gamma_1$ mode also has an unusually high quantum yield ($\phi$) for its excitation. As shown in the $g(V)$ spectrum taken from the center of a $\mathrm{T}_3$ cluster [Fig.~\ref{fig-inelastic-analysis}(a)], $g$ nearly doubles as the energy exceeds the peak energy of $\gamma_1$, indicating that almost half of the electrons that have sufficient energy to excite the transition undergo inelastic tunneling.  Taking a ratio of the two $g$ values (one below $E_{\gamma_1}$ and one above) leads to an estimate of $\phi$ of $\approx 0.5$, with similar step heights for all inelastic spectra of the other types of clusters (see Supplemental Note III and Supplemental Fig. S6). We expect that the observed high quantum yield is a consequence of the two-dimensional Fermi surface of the Pd layer with no states near the $\Gamma$ point, so that phonon-assisted tunneling as observed, e.g., in graphene \cite{zhang_giant_2008} results in high cross-sections for inelastic tunneling.

The lifetime of the $\gamma_1$ mode determined from its apparent energy width ($\sim 0.3~\mathrm{ps}$) [Fig.~\ref{fig-inelastic-analysis}(a), and see Supplemental Note IV and Supplemental Fig.~S7] is of a similar order of magnitude to those of the internal stretching mode of \ce{CO} molecules chemisorbed on \ce{Pd(100)} and on \ce{Pt(111)} \cite{persson_vibrational_1980}, and appreciably shorter than those physisorbed on \ce{Au(111)} ($49\pm3~\mathrm{ps}$) \cite{kumar_vibrational_2019}.

\subsection{Determination of vibrational lifetime}
A determination of the lifetime of the vibrational state from its linewidth is not possible, because the width is comparable to the intrinsic resolution limit given by thermal broadening. We have, however, been able to assess the lifetime through driving population of the excited state with tunneling electrons. To this end, we studied the dependence of the intensity of the excited state on the tunneling current. For high current and hence excitation rate, the ground state is expected to become depopulated, resulting in a decrease in the intensity at its fundamental frequency. In Fig.~\ref{fig-inelastic-analysis}(b-c) we show an analysis of the dependence of the intensity of the $\gamma_1$ mode on the tunneling current. We have performed detailed $g'(V)$ spectroscopy measurements with increasing set-point current $I_\mathrm{s}$ on five different cluster types: $\mathrm{T_1}$, $\mathrm{T_3}$ to $\mathrm{T_5}$, and $\mathrm{T_7}$.  Shown in Fig.~\ref{fig-inelastic-analysis}(b) are the $g'(V)$ spectra recorded at increasing $I_\mathrm{s}$ from the central position of the clusters. To facilitate a meaningful comparison between the spectra recorded at different $I_\mathrm{s}$, we normalize each $g'(V)$ spectrum by the value of the tunneling current at the peak energy of the $\gamma_1$ mode ($I_p$). Through numerical fitting to the experimental data, we have determined the normalized peak height $h_\mathrm{norm}$, which reflects the excitation probability per electron of the $\gamma_1$ mode at different $I_\mathrm{s}$. Shown in Fig.~\ref{fig-inelastic-analysis}(c), our results reveal a monotonic decrease in $h_\mathrm{norm}$ of the $\gamma_1$ mode with increasing $I_\mathrm{s}$, suggesting that ground state depopulation of $\gamma_1$ may already occur when $I_\mathrm{s}$ exceeds $50~\mathrm{pA}$. Such depopulation is also observed in other cluster types.

Although all plots show a monotonic decrease in $h_\mathrm{norm}$ with increasing $I_\mathrm{s}$, the decrease differs qualitatively for different types of clusters, which we rationalize by the different lifetimes of the excited state associated with the $\gamma_1$ mode in each type of cluster.  To verify this, we performed a numerical fitting to the $h_{\mathrm{norm}}$ versus $I_{\mathrm{s}}$ data [Fig.~\ref{fig-inelastic-analysis}(c)] using a model through which the lifetime of the $\gamma_1$ mode can be determined for each type of cluster. In the model [Fig.~\ref{fig-inelastic-analysis}(d)], we assume that the excitation transition can only take place between the ground and the first excited state, and only when the ground state is populated. With this assumption, the time-average probability that the mode is in the ground state has the form $N_g = \frac{R}{R+1}$ with $R = \kappa+\left(\Gamma\tau\right)^{-1}$, where $\Gamma=\frac{\phi I}{e}$ is the excitation rate, $\phi$ the quantum yield of the excitation, $\tau$ the spontaneous relaxation lifetime of the mode, $I$ the tunneling current, $\kappa$ the ratio of stimulated de-excitation to excitation rates ($\Gamma_\mathrm{down} =\kappa \Gamma_\mathrm{up}$), and $e$ the electron charge. See Supplemental Note V for the derivation of the fit function of the two-level model. 

To compare this behavior with the one observed experimentally, we fit the function $h(I) = h_0 N_g$ to the change in amplitude $h(I)$, where $h_0$ is a proportionality factor. With this, our fit to the experimental data (Fig.~\ref{fig-inelastic-analysis}) reveals 
estimated lifetimes (expressed in $\phi \tau$) ranging between $11.3~\mathrm{ps}$ and $519~\mathrm{ps}$, and $\kappa$ (which accounts for stimulated deexcitation) between $0.04$ and $0.69$ for different cluster types. We could not obtain a reasonable fitted value for the lifetime of the cluster $\mathrm{T}_5$, and this is probably due to the lack of sufficient data points available for fitting.

To take into account possible high order excitations that may also take place in inelastic excitations, we performed data fitting using an extended $n$-level model with increasing $n$, examining any $n$-dependent changes of the fitted parameters. The $n$-level model has its expression of the probability that the system is in the ground state as $N_g(n)=\frac{R^{n-1}}{\sum_{j = 1}^{n}R^{j-1}}$ (with $R=\kappa+\frac{1}{\Gamma\tau}$). Our results (Supplemental Fig.~S8) show that increasing $n$ does not lead to any significant change in the fitted lifetime ($\phi\tau$). However, it causes a continuous increase in $\kappa$, which finally increases by unity as $n\rightarrow\infty$. See Supplemental Note VI and Supplemental Fig.~S8 for more details.

Back to our discussion of the results based on the two-level model, assuming that the quantum yield ($\phi$) is the same for all types of clusters, we find that the lifetimes of the excited states of the clusters $\mathrm{T}_1$ and $\mathrm{T}_7$ are on average an order of magnitude longer than those of other cluster types. This likely arises from their perfect geometries, which amplify resonant excitations that may already occur within the clusters as a result of spatial confinement \cite{yim_adsorbate-induced_2025}, in turn inhibit dissipation, resulting in a longer lifetime.  Evidence of the population of the excited state is the more protruding appearance of the central atom of each of the two cluster types at higher tunneling currents (see Supplemental Note VII and Supplemental Fig.~S9).

Another potential factor contributing to the longer lifetimes in clusters $\mathrm{T}_1$ and $\mathrm{T}_7$ is the difference in the probed positions of the tunneling spectra: all spectra were recorded at the central positions of the clusters. For clusters $\mathrm{T}_1$ and $\mathrm{T}_7$, the central position is on top of an adsorbed H atom; for clusters $\mathrm{T}_3$ and $\mathrm{T}_4$, it is between hydrogen atoms.

In addition to the extended lifetimes, the high $\kappa$ values exhibited by clusters with higher symmetries suggest that stimulated de-excitation occurs appreciably within them. These high $\kappa$ values also explain the non-zero offset of $h_\mathrm{norm}$ at high currents, indicating a balance between stimulated excitation and de-excitation of the $\gamma_1$ mode under these conditions.

\subsection{Vibrational lifetime calculations}
The vibrational lifetimes observed from the tiling structure are appreciably longer than those of adsorbates on typical metal surfaces (from sub-picoseconds to a few picoseconds) \cite{shirhatti_observation_2018,kumar_vibrational_2019,liu_adsorbate_2006} and less than those seen on semiconducting (or insulating) surfaces ($\sim \mathrm{ns}$) \cite{PhysRevLett.64.2156,west_first-principles_2006}. We speculate that the larger lifetime compared to bulk Pd is due to the electronic decoupling of the Pd surface layer from the bulk, which hampers the effectiveness of electronic deexcitation along the stacking direction of \ce{PdCrO2}, and the significantly lower charge carrier density in \ce{PdCrO2} compared to typical metal surfaces \cite{persson_vibrational_1980}. At the same time, the existence of conduction electrons does provide a dissipation channel, reducing the lifetime compared to that of adsorbates on semiconductors and insulators.

To address this, we performed vibrational lifetime calculations of hydrogen on the Pd-terminated surface of \ce{PdCrO2}, and on the (111) surface of \ce{Pd} single crystals for comparison. The dominant modes are two in-plane bending modes and an out-of-plane stretching mode [Fig.~\ref{fig-calculation}(a)]. Our results for the lifetime [Fig.~\ref{fig-calculation}(b)] show that on the \ce{Pd}-terminated surface of \ce{PdCrO2}, the stretching mode has a shorter lifetime than the bending modes. In particular, the lifetimes of the bending modes are up to nine times longer than those of hydrogen on Pd(111), and become significantly higher when the antiferromagnetic ordering in the \ce{CrO2} layer is taken into account, indicating that electronic correlations reduce electronic screening of the vibrational process, in turn lengthening the lifetime. Our calculations therefore indicate that the extended vibrational lifetimes of hydrogen in \ce{Pd}/\ce{PdCrO2} arise primarily from the structural uniqueness of delafossite \ce{PdCrO2} compared to pure metals, and, are further increased by the correlation effects in the material that reduce electronic screening (see Supplemental Note VIII for further details, and Supplemental Fig.~S10 for calculated lifetimes of all vibrational modes present in the systems discussed above).

\subsection*{Correlation between two vibration modes}
Finally, to demonstrate a direct correlation between the in-plane and out-of-plane vibrational modes ($\gamma_1$ and $\varepsilon$), we show in {\Revise Fig.~\ref{fig-correlation}} the $g'(\mathbf{r},V)$ map slices obtained on the hydrogen tiling at their peak energies ($274~\mathrm{meV}$ for $\varepsilon$ and $42~\mathrm{meV}$ for $\gamma_1$), which show a high resemblance to each other.  Their close relationship is also reflected in the 2D histogram [{\Revise Fig.~\ref{fig-correlation}(d)}], and the simultaneous decrease in their peak intensity with increasing current (Supplemental Fig.~S11).

\section{CONCLUSIONS}
Using IETS in STM, we studied the vibrational modes within the hydrogen clusters of the non-periodic tiling formed on the Pd surface layer of delafossite oxide \ce{PdCrO2}.  Our results show that these vibrational modes have longer relaxation lifetimes, with the measured lifetimes reaching up to hundreds of picoseconds, appreciably longer than those observed on typical metal surfaces \cite{morin_vibrational_1992,germer_picosecond_1994,west_first-principles_2006,shirhatti_observation_2018,kumar_vibrational_2019}.  We attribute this phenomenon to the inherent structural and electronic anisotropies of the delafossite oxide, weak coupling between the surface layer and the bulk, and correlation effects in \ce{PdCrO2}, which collectively lead to effective shielding of the vibrational modes from the surroundings, resulting in inefficient dissipation.  Furthermore, our findings provide insight into the exceptional electrocatalytic activity observed during the hydrogen evolution reaction on the \ce{Pd} surface of \ce{PdCoO2} \cite{li_situ_2019,podjaski_rational_2020}, which surpasses that of nanostructured palladium \cite{moumaneix_interactions_2023,metzroth_accelerating_2021,darmadi_high-performance_2020}.

\section{METHODS}
\subsection{Scanning tunneling microscopy/spectroscopy (STM/S)}
The STM measurements were performed using a home-built low temperature STM machine that operates at a base temperature of $1.8\,\mathrm{K}$ \cite{white_stiff_2011}.  \ce{Pt/Ir} tips were used and conditioned by field emission performed on a \ce{Au} target before use.  Differential conductance $g(V)$ ($\mathrm{d}I/\mathrm{d}V$) and inelastic electron tunneling $g'(V)$ ($\mathrm{d}^2I/\mathrm{d}V^2$) spectra and spectroscopic maps were recorded using a standard lock-in technique, with the frequency of the bias modulation set at $413~\mathrm{Hz}$.   To ensure sufficient data quality in the set-point current ($I_s$) dependent $g'(V)$ spectroscopy measurements, in particular at low currents, the spectrum at each $I_s$ was recorded $n_\mathrm{rep}$ times so that $I_s\times n_\mathrm{rep} \geq 2\times 10^{3}\,\mathrm{pA}$.
All reported data were obtained at a sample temperature of $4.2\,\mathrm{K}$ unless otherwise stated.  For STM/S measurements, the clean sample surfaces were prepared by cleaving the samples in situ at a sample temperature of $\sim 20\,\mathrm{K}$.

\subsection{Sample growth}
Single-crystal samples of \ce{PdCrO2} were grown by the \ce{NaCl}-flux method as reported in Ref.~\onlinecite{takatsu_single_2010}. First, polycrystalline \ce{PdCrO2} powder was prepared from the following reaction at $960~^\circ\mathrm{C}$ for four days in an evacuated quartz ampoule: 
\begin{equation}
    \ce{2LiCrO2 + Pd + PdCl2 -> 2PdCrO2 + 2LiCl}.
\end{equation} 
The powder obtained was washed with water and aqua regia to remove \ce{LiCl}. Polycrystalline \ce{PdCrO2} and \ce{NaCl} were mixed in a molar ratio of 1:10. The mixture in a sealed quartz tube was then heated at $900~^\circ\mathrm{C}$ and slowly cooled to $750~^\circ\mathrm{C}$. \ce{PdCrO2} single crystals were harvested after dissolving the \ce{NaCl} flux with water.

\subsection{Theoretical calculations}
\textit{Ab initio} density functional theory (DFT) calculations were performed with the Fritz Haber Institute Ab Initio Molecular Simulation (FHI-aims) numeric atomic orbital software package \cite{blum_ab_2009}, which is an all-electron, full potential code. The general gradient approximation exchange-correlation functional  of Perdew-Burke-Ernzerhof (PBE) was used \cite{perdew_generalized_1996}. Calculations were performed using a ‘light’ basis set of the 2010 release, where self-consistent field cycle convergence reached when the sum of eigenvalues and the change in charge density were below $10^{-6}~e/a_0^3$ and $10^{-6}~\mathrm{eV}$, respectively. The calculations were spin-polarized to account for the antiferromagnetic ordering of the \ce{CrO2} layers for \ce{PdCrO2} and spin-restricted for Pd(111). The zeroth order regular approximations was used for relativistic treatments \cite{van_lenthe_relativistic_1994}.

The structures of \ce{PdCrO2} and Pd(111) were built and manipulated using the Atomic Simulation Environment (ASE) Python library \cite{hjorth_larsen_atomic_2017}. The unit cell structure model of \ce{PdCrO2} was subjected to full geometry and atomic optimization, \textit{i.e.}, lattice parameters and atom positions, using the Broyden-Fletcher-Goldfarb-Shanno (BFGS) algorithm with convergence reached when all forces on all atoms were less than $0.01\,\mathrm{eV}$\AA$^{-1}$ \cite{broyden_convergence_1970, fletcher_new_1970, goldfarb_family_1970, shanno_conditioning_1970}, which yielded lattice parameters of \textit{a},\textit{b} = 2.97 \AA\ and \textit{c} = 18.19 \AA\ commensurate with experiment \cite{shannon_chemistry_1971, takatsu_critical_2009}. Symmetric, Pd-terminated $2\times2\times1$ supercells were built from the unit cell of \ce{PdCrO2}, with a slab thickness of 7 layers and a vacuum of 15 \AA\ in the $z$ direction, which were subject to geometry optimization only, using the BFGS algorithm and the same convergence criteria as the unit cell. For Pd(111) $4\times4\times6$ supercells were built, with a slab thickness of 6 layers and a vacuum of 15 \AA\ in the $z$ direction, which were also subject to geometry optimization using the BFGS algorithm with the same criteria as \ce{PdCrO2}.  A converged Monkhorst-Pack $k$-point sampling grid of $4\times4\times1$ was used on the slab structures of PdCrO$_2$ and Pd(111) \cite{monkhorst_special_1976}. To account for the antiferromagnetic ordering of the \ce{CrO2} layer a striped ordering was adopted, where non-collinear ordering was ignored for simplicity. 

For surface reactions, the adsorption energies, $E_\mathrm{ads}$,  measure the energy between the surface and the reactant, and are calculated through comparison of the energy of the optimized gas-phase adsorbate, $E_\mathrm{A}$, optimized surface, $E_\mathrm{S}$, and the combined system $E_\mathrm{A-S}$, via: 

\begin{equation}
E_\mathrm{ads}= E_\mathrm{A-S} - (E_\mathrm{A} + E_\mathrm{S})
\end{equation}

where a negative value indicates favorable adsorption. 

Vibrational lifetime analysis was performed using the implementation in FHI-aims \cite{box_ab_2023} through the evaluation of the electronic friction which was computed from the matrix elements of electron-phonon coupling as represented in the equation below: 

\begin{equation}
\gamma_\alpha = \hbar \left[ \tilde{u}_\alpha^{T} \wedge (\hbar \omega_\alpha)\, \tilde{u}_\alpha \right]
\end{equation}

where $\gamma_\alpha$ is the vibrational line-width of the phonon,  $\Lambda$ the electronic friction tensor, $\hbar$ the reduced Planck's constant and $\tilde{u}_\alpha$ the mass-weighted vibrational eigenmodes at phonon frequency $\omega_\alpha$. The vibrational lifetime is related to the phonon linewidth via: 

\begin{equation}
\tau_\alpha = \hbar / \gamma_\alpha
\end{equation}

\section*{ACKNOWLEDGEMENTS}
C.G. and O.T.B. thank Reinhard Maurer and Connor Box for help with code for computing vibrational lifetimes. C.M.Y. acknowledges support from the Ministry of Science and Technology of China (2022YFA1402702, 2021ZD0302700) and National Natural Science Foundation of China (12574524). C.M.Y. and P.W. acknowledge support from EPSRC (EP/S005005/1). C.G. was supported by the EPSRC through a New Investigator Award [grant number UKRI132]. We are grateful to the UK Materials and Molecular Modelling Hub for computational resources, which is partially funded by EPSRC (EP/T022213/1, EP/W032260/1 and EP/P020194/1)\\

\section*{DATA AVAILABILITY}
The data that support the findings of this article are openly available [\onlinecite{yim_raw_2026}].\\

\bibliographystyle{unsrt}

\clearpage
\begin{figure}
    \centering
    \caption{(a) Side view schematic model of the delafossite oxide \ce{PdCrO2} with a Pd terminated surface exposed and that adsorbed with hydrogen (pink colored spheres). (b) Top view model showing that the tiling structure, formed as a result of hydrogen adsorption, comprises tightly packed hexagonal clusters of hydrogen of different sizes and shapes. (c)-(d) STM topographic images of (c) the pristine Pd-terminated surface of \ce{PdCrO2} and (d) that adsorbed with hydrogen, resulting in a tiling formed on the surface.   Image size: $(7.5\times12)~\mathrm{nm}^2$. Scan parameters $(V, I)$: (b) $(200~\mathrm{mV}, 50~\mathrm{pA})$, (c)  $(52.5~\mathrm{mV}$, $60~\mathrm{pA})$.
    }
    \label{fig-intro}
\end{figure}

\begin{figure}
    \centering
    \caption{(a) $g(V)$ spectrum recorded at the center of one cluster of the non-periodic tiling. ($V_\mathrm{s}=500~\mathrm{mV}$, $I_\mathrm{s}=10~\mathrm{nA}$; $V_\mathrm{m}=5~\mathrm{mV}$). (b) Corresponding $g'(V)$ and $-g'(-V)$ spectra recorded simultaneously with the $g(V)$ spectrum in (a).  Six inelastic features were identified through comparison between $g'(V)$ and $-g'(-V)$: $\alpha$ mode at $3~\mathrm{meV}$, $\beta$ at $15~\mathrm{meV}$, $\gamma_1$ and $\gamma_2$ at $42~\mathrm{meV}$ and $84~\mathrm{meV}$, $\delta$ at $59~\mathrm{meV}$, and $\varepsilon$ at $274~\mathrm{meV}$. (c) STM topographic image of the non-periodic tiling [$V=-200~\mathrm{mV}$, {$I=60~\mathrm{pA}$}; image size: $(8~\mathrm{nm})^2$]. (d)-(e) $g' (\mathbf{r},V)$ map slices on (d) and off (e) the $\gamma_1$ peak energy at energies of $42.5~\mathrm{meV}$ and $35~\mathrm{meV}$ recorded simultaneously with (c) ($V_\mathrm{s}=-200~\mathrm{mV}$, {$I_\mathrm{s}=5~\mathrm{nA}$ }; $V_\mathrm{m}=2.5~\mathrm{mV}$).  (f) Averaged $g'(V)$ spectra extracted from the central part of the clusters and at the boundary region.
    }
    \label{fig-inelastic-overall}
\end{figure}

\begin{figure}
    \centering
    \caption{(a) $g(V)$ (dark yellow) and $g'(V)$ (light yellow) point spectra taken from the center of a $\mathrm{T}_3$ cluster ($V_s=52~\mathrm{mV}$, $I_s=20~\mathrm{pA}$; $V_m=1.5~\mathrm{mV}$), showing a significant jump of $g$ crossing the peak energy of the $\gamma_1$ mode. Taking the ratio of the $g$ values on the two sides of the $g(V)$ spectrum gives a quantum yield ($\phi$) of $\approx 0.5$.  Inset: Topographic image of the $\mathrm{T}_3$ cluster, with a white dot indicating where the spectra were taken.  (b) $g' (V)/I_\mathrm{p}$ versus $V$ spectra recorded at increasing set-point current ($I_\mathrm{s}$) at the central position of each of five different cluster types. $V_\mathrm{s}=51~\mathrm{mV}$, $V_\mathrm{m}=1~\mathrm{mV}$.  $I_\mathrm{p}$ denotes the value of the tunneling current at the peak energy of the $\gamma_1$ mode.  (c) Corresponding plots of normalized peak height of the $\gamma_1$ mode as a function of $I_\mathrm{s}$ determined from fitting to experimental data using a Gaussian line-shape. Red lines are fitted curves to the experimental data using the equation derived from the vibration excitation model shown in (d), yielding the estimated excitation lifetimes expressed in $\phi\tau$ and $\kappa$, the ratio of the stimulated de-excitation to excitation rates, for different cluster types.
    Insets of (c): atomically resolved images of the studied clusters.  Solid circles indicate the positions at which the spectra were taken. (d) Schematic of the two-level vibration excitation model employed, which considers three different transitions: stimulated excitation from the ground state and stimulated and spontaneous de-excitations from the excited state, respectively.}
    \label{fig-inelastic-analysis}
\end{figure}

\begin{figure}
    \centering
    \caption{(a) Schematics of three selected modes of vibrations of three different hydrogen adsorbed systems: $\mathrm{T_1}$ hydrogen cluster on the \ce{Pd}-terminated surface of \ce{PdCrO2}, $\mathrm{T_1}$ cluster with only considering vibrations of the central H atom, and fcc-bound H on Pd(111). Green arrows indicate the displacement directions of the H atoms in each mode. Blue, red, gray, and white spheres represent Pd, O, Cr, and H, respectively. Among the schematics, The first two rows are the bending modes of vibration (low frequency $f$) and the bottom row the stretching mode (large frequency $f$). (b) (Top) Calculated lifetimes and (Bottom) frequencies of the vibrational modes shown in (a). For $\mathrm{T_1}$ cluster on \ce{Pd/PdCrO2}, results of spin-polarized (with AFM) and spin-restricted (no AFM) calculations are presented using mid- and light- blue markers, respectively.}
    \label{fig-calculation}
\end{figure}

\begin{figure}
    \centering
    \caption{(a) STM Topographic image of the non-periodic tiling. Image size: $(4.7\times4.7~\mathrm{nm})^2$. $(V,I)=-305~\mathrm{mV}, 10~\mathrm{nA}$.  (b)-(c) Simultaneously recorded $-g'(\mathbf{r},V)$ map slices at bias voltages of (b) $-273.6~\mathrm{mV}$ and (c) $-43.6~\mathrm{mV}$ [($V_\mathrm{s},I_\mathrm{s})=-305~\mathrm{mV},10~\mathrm{nA}$; $V_\mathrm{m}=7.5~\mathrm{mV}$]. (d) 2D histogram showing high correspondence between the two map slices in (b) and (c).  The horizontal and vertical dashed lines indicate the non-zero average intensity of the two map slices.}
    \label{fig-correlation}
\end{figure}

\clearpage

\begin{center}
\includegraphics[width=\columnwidth]{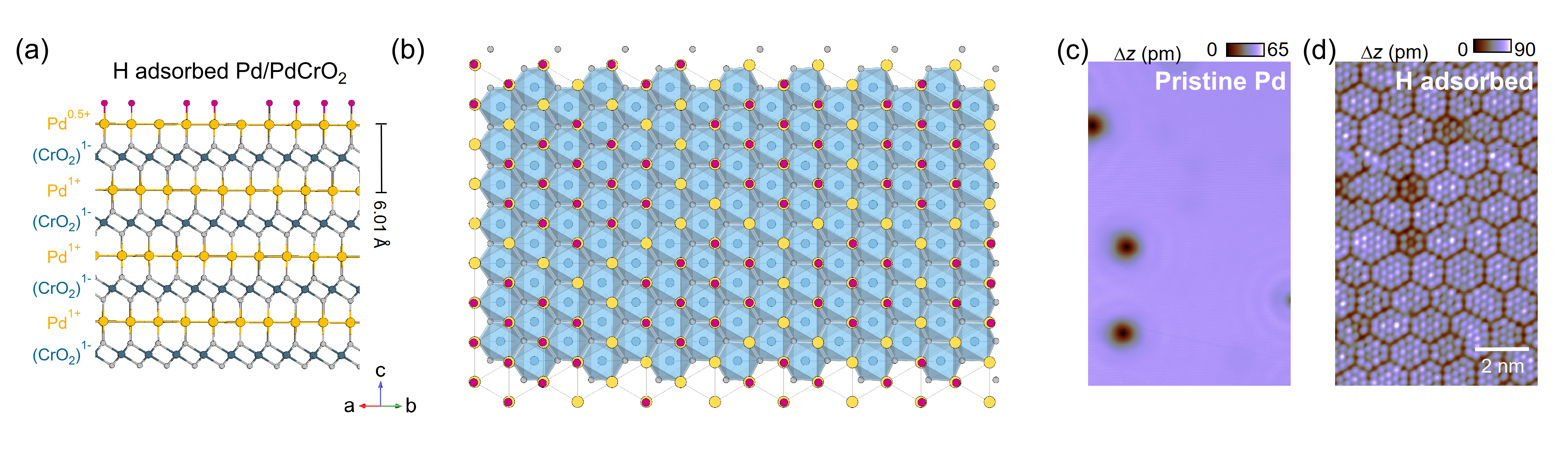}
\end{center}
\huge{FIG. 1}
\newpage
\begin{center}
\includegraphics[width=\columnwidth]{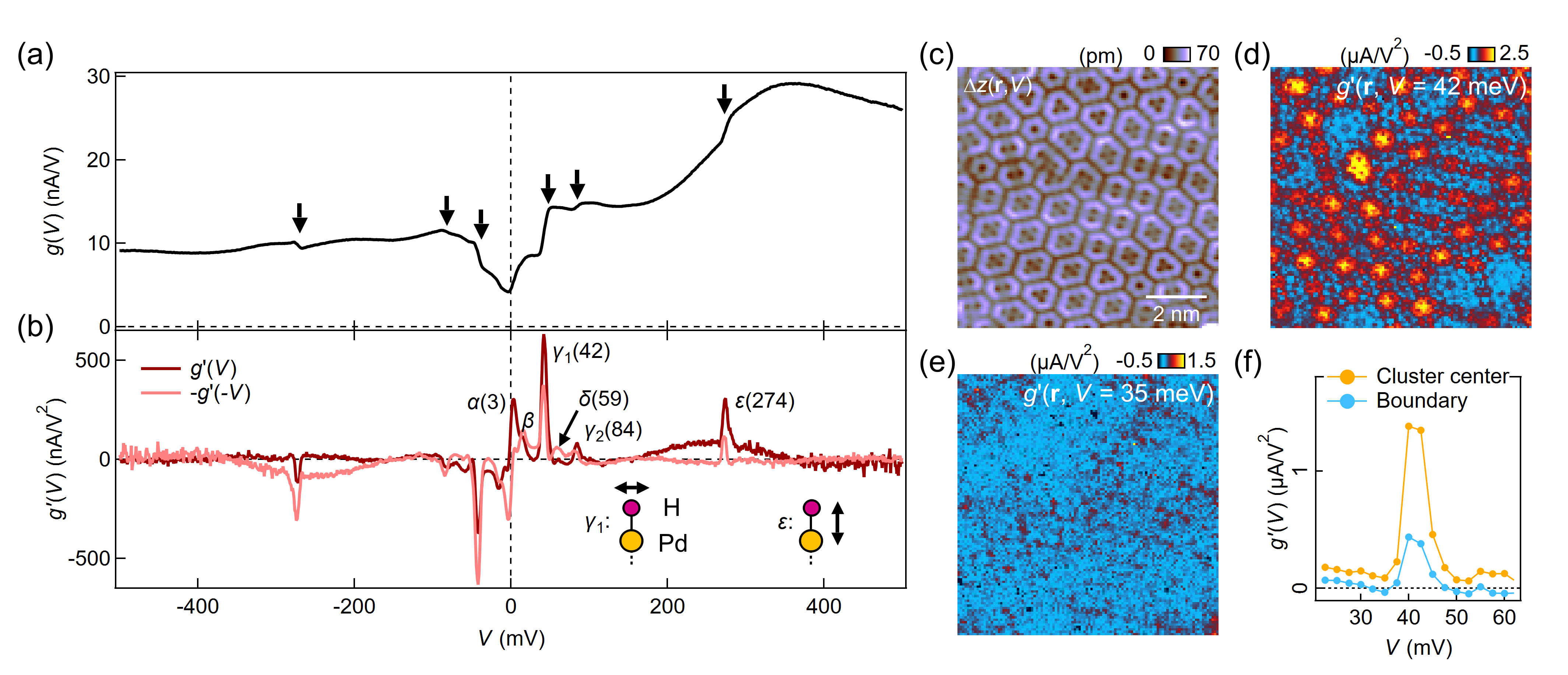}
\end{center}
\huge{FIG. 2}
\newpage
\begin{center}
\includegraphics[width=\columnwidth]{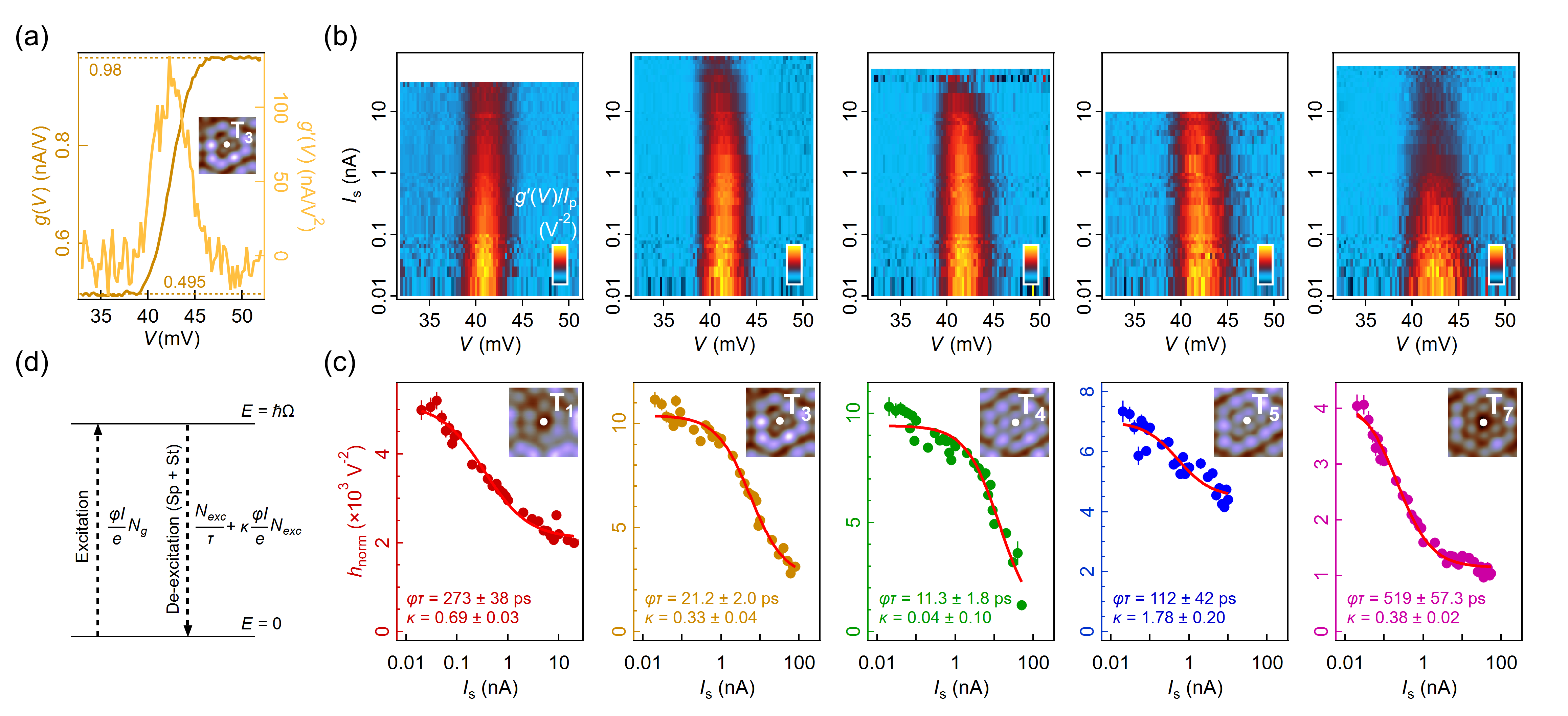}
\end{center}
\huge{FIG. 3}
\newpage
\begin{center}
\includegraphics[width=8cm]{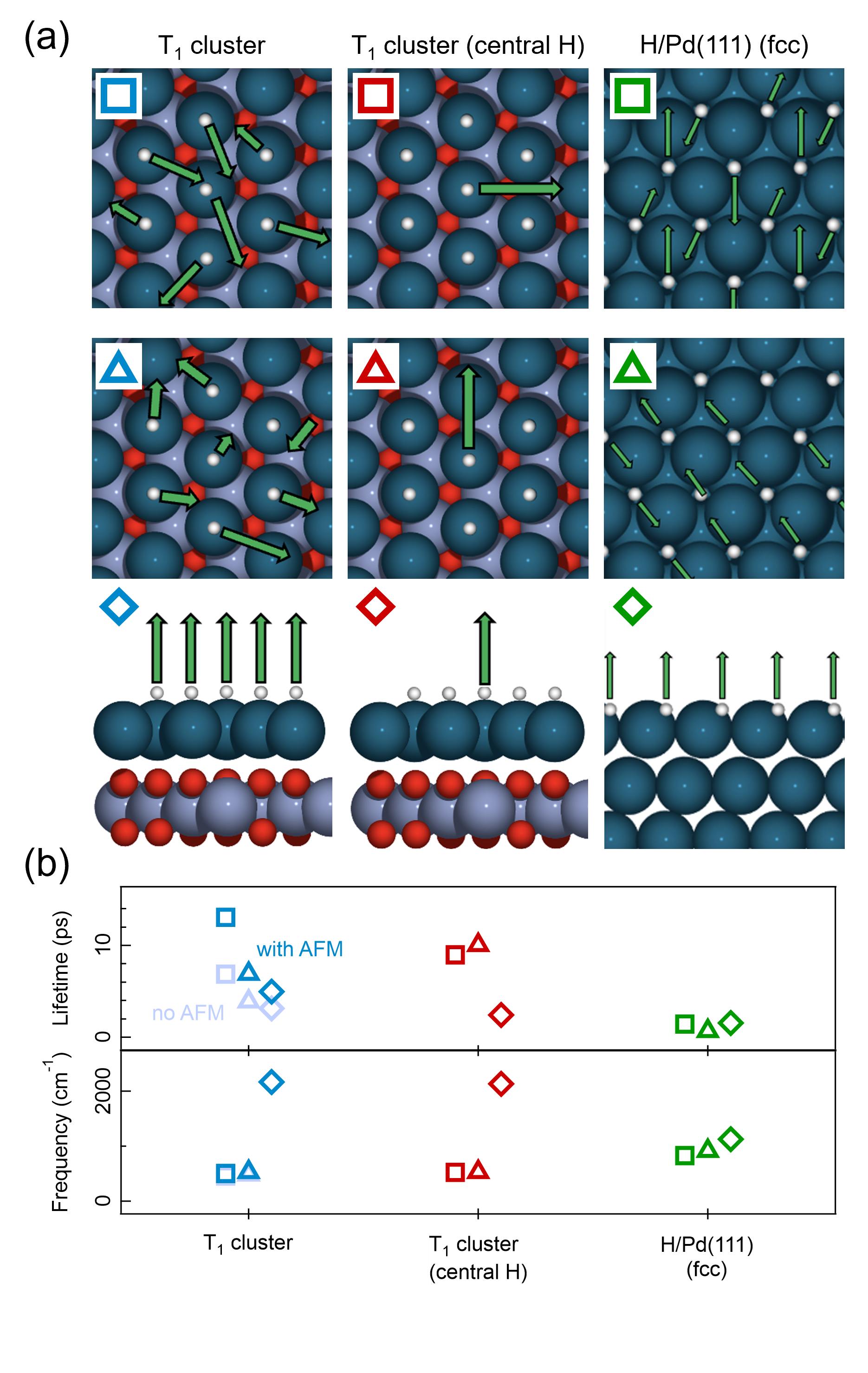}
\end{center}
\huge{FIG. 4}
\begin{center}
\includegraphics[width=8cm]{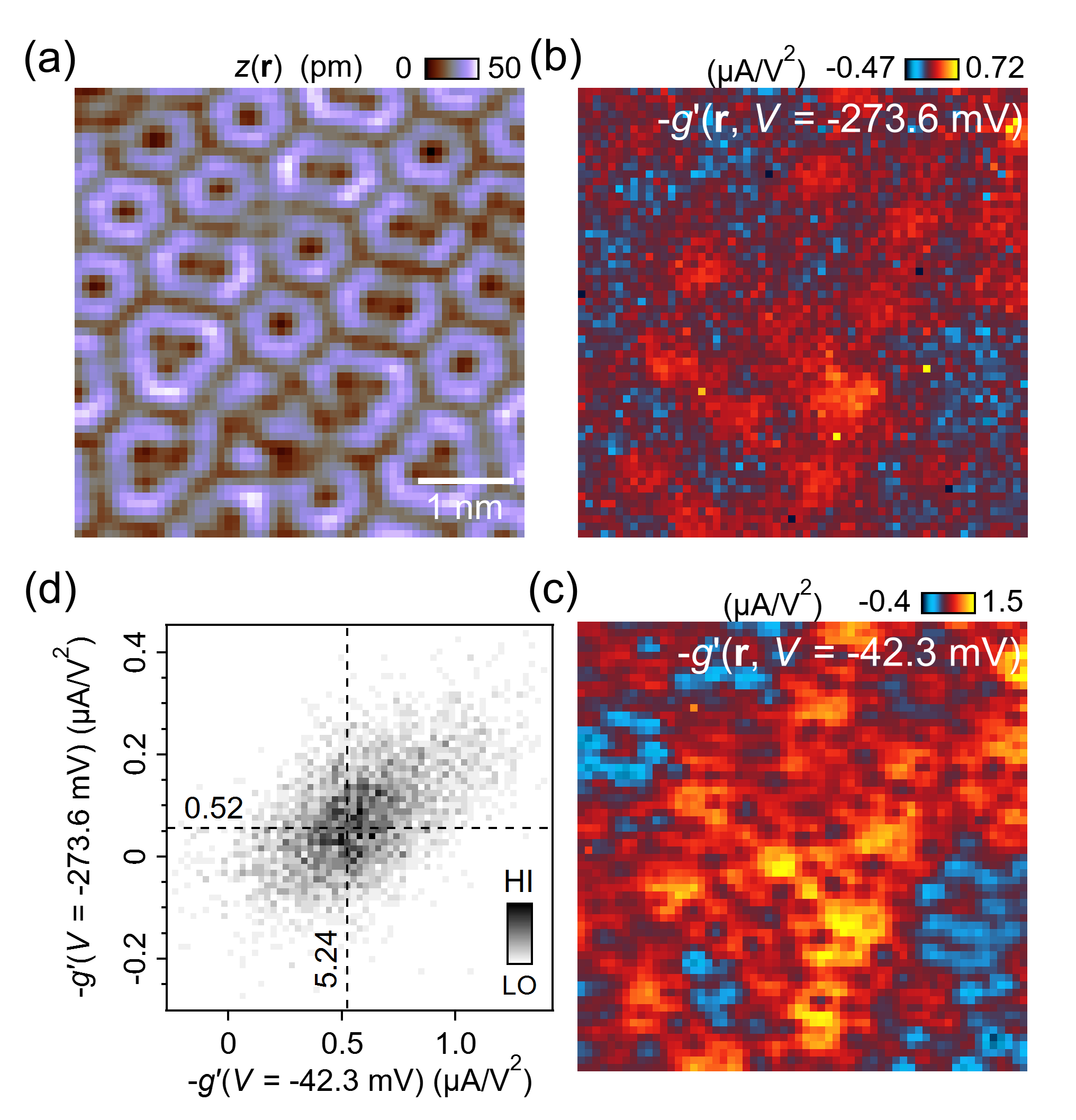}
\end{center}
\huge{FIG. 5}

\end{document}